\documentstyle[11pt,moriond,epsfig]{article}

\def \r{{\bf r}}

\def \ep{\epsilon}

\begin{document}

\vspace*{4cm}
\title{A FEW PROPERTIES OF DISORDERED CONDUCTORS}

\author{G. MONTAMBAUX}

\address{Laboratoire de Physique des Solides, Associ\'e au CNRS, \\Universit\'e Paris-Sud,
91405 - Orsay, France}

\maketitle\abstracts{
In this brief tutorial review, I show how phase coherent properties
of disordered conductors can be described in a simple and unified
way. These properties include transport properties like
weak-localization correction and universal conductance
fluctuations, and thermodynamic properties like orbital magnetism
and persistent currents. They can be related to the classical
return probability for a diffusive particle. For a network with $N$
nodes, the return probability can be related to the determinant of
a $N \times N$ "connectivity" matrix $M$ so that the magnetization
and the transport quantities can be {\it directly} written in term
of $\det M$. }

\section{Introduction}

The goal of this short review{\footnote {\it Invited talk,
International Workshop on superconductivity, magnetoresistive
materials and correlated quantum systems, Rencontres du Vietnam,
Hanoi 1999, p. 207 (Vietnam National University Press)}} is to show
that transport and thermodynamic properties of phase coherent
conductors can be described in a simple unified way in which the
quantities of interest are related to the classical return
probability for a diffusive particle. We consider weakly disordered
conductors, for which the mean free path $l_e$ is much larger than
the distance between electrons: $k_F l_e \gg 1$, where $k_F$ is the
Fermi wave vector.

Let us first consider the  effect of phase coherence on transport. The    Drude-Boltzmann   conductivity assumes that
interference between electronic waves can be neglected.
The structure of this classical
conductivity is  given by a sum of probability intensities $\sigma_{cl} \propto \sum_{j} |A_j|^2$ where $A_j$
represents some amplitude related to a diffusion process. However, one knows that in quantum mechanics one must add
amplitudes instead of intensities. Thus, the structure of the
conductivity has to be $\sigma \propto  \sum_{i,j} A_i A_j^*  $.
Since the interference terms, of the form $A_i A_j^*$, have random phases, they cancel in average so that the conductivity
reduces to its classical value given by the diagonal terms in the sum.
However, there is a class of  contributions which may not cancel in average. They correspond to diffusive
trajectories which
form {\it closed loops}.
Such a loop can be traveled in clockwise or
 anti-clockwise directions. If there is time reversal symmetry, both trajectories, $j$ and its time-reversed
$j^T$, have the same action, so that they interfere constructively.
As a result, in addition to the classical average conductivity, there is a
correction of the form  $\Delta \sigma \propto \langle\sum'_{j} A_j A^*_{j^T}\rangle$
where the sum extends over the closed trajectories.
The sign of the correction is negative
 because the trajectories $j$ and $j^T$ have opposite
momenta. The
conductivity is thus {\it reduced} and the correction is called {\it weak-localization correction} \cite{wl,Chakravarty86}.
This is a
phase
coherent effect because it implies that the time reversed trajectories have the same action. This phase coherence is broken
by inelastic events due to the coupling to other degrees of freedom or due to electron-electron interactions. Such
coherence breakdown is temperature dependent and can be phenomenologically described by a phase coherence length
$L_\phi$:
trajectories larger than $L_\phi$ cannot contribute to the weak-localization corrections.
The effect of a magnetic field which breaks time-reversal symmetry is to destroy this phase-coherent contribution to the
conductance, leading to a {\it negative magnetoresistance} (in the absence of spin-orbit scattering) which is one of  the most spectacular signature of phase
coherence.

Another important signature
 of the coherent nature of quantum transport is
the phenomenon of {\it Universal Conductance Fluctuations}
\cite{ucf,Altshuler86}. When a physical parameter is varied, such
as the Fermi energy, the magnetic field or the disorder
configuration, the conductance fluctuates around its average value.
These fluctuations are reproducible and are the signature of the
interference pattern associated to a given impurity configuration.
The width of the distribution is universal and of the order of
$e^2/h$ The structure of the variance of the conductivity is
$\delta \sigma^2 \propto \sum_{i,j,k,l} [\langle A_i A_j^*A_k
A_l^*\rangle -\langle A_i A_j^*\rangle\langle A_k A_l^*\rangle]$.
Correlation terms imply pairs of diffusive trajectories.

These transport properties have been extensively studied in the 70'
and 80'. More recently, a wide interest emerged in the experimental
studies of the equilibrium properties. Among them the search for
the persistent current of an isolated mesoscopic metallic ring
pierced by a magnetic Aharonov-Bohm flux $\phi$. More generally,
one is interested in the magnetization $M$ of a phase coherent
system. In the thermodynamic limit, its average value is nothing
but the Landau magnetization. However the fluctuations of the
magnetization depend on phase coherence \cite{Shapiro,Oh91}.
Moreover, it is known that considering electron-electron
interactions leads to an additional phase coherent contribution to
the magnetization which can become larger than the Landau
magnetization \cite{AAZ,Oh91}. In the geometry of a ring, this
magnetization corresponds to a current $I$ flowing along the ring.
This persistent current has been observed in single or arrays of
isolated mesoscopic rings \cite{cpexp}.

In this paper, I briefly review the derivation of these different transport and thermodynamic quantities.
I show that they are related to the return probability for a diffusive particle. I treat a few examples and I present a
formalism
to calculate simply all these quantities on networks of any geometry.

\section{Phase coherence and Diffusion in a disordered medium}

Technically, the quantities of interest are either response functions or equilibrium quantities, which can be written in
terms of
products of two single particle Green functions $G_\ep(\r,\r')$, solutions of the Schr\"odinger equation for a particle in
a disordered potential $V(\r)$:
\begin{equation}
[\ep + {\hbar^2 \over 2 m} \Delta - V(\r) ] G_\ep(\r,\r')
=\delta(\r-\r')
\end{equation}
By definition, the probability for a particle to evolve from one point $\r$ to another $\r'$ is also related to the product
of two propagators. After disorder averaging, one can show that in the limit $k_F l_e \gg 1$ and for slow variations, the
probability
$P(\r,\r',\omega)$ defined as \cite{Chakravarty86}:

\begin{equation}
P({\bf r},{\bf r}',\omega) = {1  \over 2 \pi \rho_0}
 \overline{G^R_\ep({\bf r},{\bf r}')
 G^A_{\ep-\omega}({\bf r}',{\bf r})}
\label{proba}
\end{equation}
where $G^R$ and $G^A$ are the retarded and advanced Green functions, is the solution of a classical diffusion equation:
\begin{equation}  \label{diff1}
[ - i\omega - D \Delta ]P_{cl}(\r,\r',\omega)= \delta(\r-\r')
\end{equation} where $D$ is the diffusion coefficient: $D=v_F
l_e/d$. $v_F$ is the Fermi velocity and $d$ is the space
dimensionality.

$G_\ep(\r,\r')$ describes the electronic propagation from $\r$ to $\r'$ and  the probability $P(\r,\r',\omega)$ is the sum
of the
contributions of pair of trajectoires, each trajectory being characterized by an amplitude and a phase proportional to its
action. For most of these pairs, the phase difference is large so that in average, their contribution cancel. Then the
 probability is essentially given by the sum of the intensities corresponding to the modulus square of the contribution of
trajetories.
 However, when $\r \simeq \r'$, the trajectories form {\it closed loops}.  Both a trajectory and its time-reversed have
the same action, so that they interfere
constructively.
As a result, when there is time reversal symmetry, the return probability is {\it doubled} compared to its classical value.
This is a phase
coherent effect because
only trajectories of size smaller than the phase coherence length $L_\phi$ will contribute to
this additional contribution.

As a result, the return probability  has two components,
 a purely classical one and an interference term which results
from interferences between pairs of time-reversed trajectories. In the
diagrammatic picture, they are related respectively to the diffuson and Cooperon diagrams \cite{wl}.
The interference term, $P_{int}(\r,\r,\omega)$, is field dependent
and, in the weak-field limit $\omega_c \tau_e \ll 1$, it is
solution of the diffusion equation  \cite{wl,Chakravarty86}:
\begin{equation} \label{diff2}
[\gamma - i\omega -  D({\bf \nabla} + {2 i e {\bf A}\over \hbar c
})^2]P_{int}(\r,\r',\omega)=  \delta(\r-\r')  \end{equation}
whose solution has to be taken at $\r'=\r$. The scattering rate $\gamma = 1 /\tau_\phi= D
/L_\phi^2$ describes the breaking of phase coherence.
$L_\phi$ is the phase coherence length and $\tau_\phi$ is the phase coherence time. $\gamma$ has to be compared to the
Thouless rate
$1/\tau_D=D/L^2$ where $\tau_D$ is the diffusion time, typical time to
diffuse through the system of linear size $L$.

Finally, let us define the space integrated (dimensionless) return probability:
$$P(t)=\int P(\r,\r,t) d\r $$
In a magnetic field, it is the sum of a classical term and an interference term:
$$P(t,B)=P_{cl}(t)+P_{int}(t,B)$$
 and can be written as
\begin{equation}
P(t)=\sum_{n}( e^{- E_n^{(cl)} t}+e^{- E_n^{(int)} t})
\label{partition}
\end{equation}
where the $E_n^{(cl)}$ and the $E_n^{(int)}$ are the eigenvalues of the  equations associated respectively to the equations
(\ref{diff1}) and (\ref{diff2}).

\begin{equation}
-D \Delta_\r \psi_n(\r) = E_n^{(cl)} \psi_n(\r)\ \ \ \ \mbox{and} \ \  \ \ - D (\nabla_\r + {2 i e {\bf A} \over \hbar c})^2
\psi_n(\r) =
E_n^{(int)} \psi_n(\r)
\label{schrodinger}
\end{equation}
Note that these equations have the same structure as a
Schr\"odinger equation, with the substitution $ {\hbar \over 2 m}
\rightarrow D$ and $e \rightarrow 2 e$ \cite{Boundaries}. Let us
now turn to the study of the physical quantities of interest and
write them as functions of $P(t)$.

\section{Weak localization}\label{WeakLoc}

The first quantum correction to the classical Drude-Boltzmann
conductivity is  called the {\it weak-localization} correction
 \cite{wl,Chakravarty86}. Linear
response theory shows that
 the d.c. $T=0K$ average conductivity  can be written as:
\begin{equation}
\langle \sigma \rangle = - { s e^2 \hbar^3 \over 2 \pi m^2 V } \int \int d{\bf
r} d{\bf r'}   \langle \partial_x G^{R}_\ep({\bf r},{\bf r'})
  \partial_{x'}  G^{A}_\ep({\bf r'},{\bf r})\rangle
\label{Kubo}
\end{equation}$s=2$ is the spin degeneracy. Eq. (\ref{Kubo}) has a structure very similar to the return probability (\ref{proba}).
As we have seen in the introduction, the amplitude of the correction is proportional to the total number of
loops of time reversed trajectories. The number of loops of length $v_F t$ being proportional
 to the return probability
$P(t)$, one deduces that the total correction is proportional to the time integrated interference part of the return
probability \cite{wl}. More precisely, it can be shown that:
\begin{equation}
\begin{array}{|c|}
\hline
\\
\displaystyle
\Delta \sigma =- \sigma  {\Delta  \over \pi \hbar}
\int_0^\infty P_{int}(t) [e^{-\gamma t}-e^{-t/\tau_e}] dt\\
\\
\hline
\end{array}
 \label{WL} \end{equation}
$\Delta$ is the interlevel spacing (for one spin direction), and $\sigma=s e^2 D \rho_0$ where $\rho_0$ is the density of states (DOS) for one
spin direction: $\rho_0=1/\Delta \Omega$, $\Omega$ being the volume of the system. The contribution of the return
probability is integrated between $\tau_e$, the smallest time
for diffusion, and $\tau_\phi=1/\gamma$, the time after which the electron looses
phase coherence.
A magnetic field
or a Aharonov-Bohm flux,  by breaking the time-reversal
symmetry, destroys the weak-localization correction.

\section{Universal Conductance Fluctuations}\label{UnivCF}

An important signature of the coherent nature of quantum transport is
the phenomenon of Universal Conductance
Fluctuations  \cite{ucf}.
The averaged square of the conductance contains terms of the form
 $\langle G^R(\r_1 ,\r'_1)G^A(\r'_1 ,\r_1)
G^R(\r_2 ,\r'_2)G^A(\r'_2 ,\r_2)\rangle$ (for clarity we omitted the
gradients). After averaging, two contractions are possible:
$\r'_1=\r_1 , \r'_2=\r_2$ and  $\r_2=\r_1 , \r'_2=\r'_1$.
The first term a) is proportional to $\int
\langle G^R(\r_1,\r_1,t)G^A(\r_2,\r_2,t)\rangle dt
d\r_1 d\r_2$ where    $\r_1$ and $\r_2$
belong to the same orbit of length $v_F t$.
Therefore integration on $\r_2$ gives a factor proportional to $v_F t$ and
the corresponding contribution to the conductance fluctuation has the form:
\begin{equation} \label{UCF1}
{\langle \delta g^2 \rangle  \over \langle g \rangle^2}  \propto
\int_0^\infty t P(t) dt
\end{equation}

 The second term b) is proportional
to
 $\int P(\r,\r',t)P(\r',\r,\tau)dt d\tau d\r d\r'$. It
 can be also rewritten in the form $\int t P(t) dt$. It describes the
contribution of the fluctuations of the diffusion coefficient to the
conductance fluctuations \cite{ucf}.
Finally, adding these two contributions, one can show that\footnote{The lower bound of the time integrals is
actually the mean collision time $\tau_e$ above which diffusion takes
place.}:
\begin{equation}
\begin{array}{|c|}
\hline
\\
\displaystyle
\langle \delta \sigma^2 \rangle  =  \sigma^2 {3  \Delta^2 \over \beta \pi^2 \hbar^2} \int_0^\infty t \ P(t)
e^{-\gamma t}d
t\\
\\
\hline
\end{array}
 \label{UCF} \end{equation}
where $\beta=1$ if there is time reversal symmetry  and $\beta=2$ in the absence of such symmetry.

The conductivity $\sigma$ is proportional to the diffusion coefficient $D$. Since the integral (\ref{UCF})
scales as the square of the characteristic time $\tau_D \propto  1/  D $, one then concludes
 that the fluctuations
are {\it universal}, in the sense that they do not depend on the disorder strengh. Good and bad conductors have different
conductivities. But the fluctuation of the conductivity when some parameter is varied (gate voltage, magnetic field or
impurity configuration) is
universal.

\section{Orbital magnetism}\label{OM}

I now consider the magnetic response of a disordered electron gas and more specifically the geometries of a
2D gas, of a quasi-1D ring.
Since the system is disordered, the magnetic response has to be defined by its distribution. Like for the conductance, I
will consider the average and the fluctuation of the magnetization.

\subsection{Average Magnetization}

{\it Non interacting particles}
\medskip

let us first neglect the interactions between electrons, as we have done for the calculation of the conductance.
The magnetization $M(B)$ is the derivative
of the grand potential $A$ with respect to the  magnetic field $B$: $M =
-{\partial A \over \partial B}$.
Introducing the
field dependent DOS for one spin direction,   $\rho(\ep,B)$,
the magnetization
can be written at zero temperature as (taking the spin into account):
\begin{equation} \label{Isum}
M = - 2{\partial \over \partial B} \int_{-\ep_F}^0 \ep \rho(\ep,B)
d\ep \end{equation}
The origin of energies is taken at the Fermi energy.
The average magnetization is thus related to the field dependence of the
average density of states $\langle \rho(\ep,B) \rangle$. In a bulk system,
this leads to the Landau diamagnetism. In the thin ring geometry where no
field penetrates the conductor, the DOS only depends on the
Aharonov-Bohm flux. Its average is flux independent
 because the flux modifies only the phase factors of the propagator
which cancel in average \cite{canonical}.
\bigskip

\noindent
{\it Contribution due to interactions}
\medskip

In the Hartree-Fock approximation,
the total energy $E_T$ is now
\begin{eqnarray} E_T & = & E_T^0 +
 {1 \over 2}\sum_{i,j} \int U(\r-\r') |\psi_j(\r')|^2 |\psi_i(\r)|^2  d\r
d\r' \nonumber  \\ & & - {1 \over 2} \sum_{i,j}
\delta_{\sigma_i \sigma_j} \int U(\r-\r')  \psi_j^{\ast}(\r') \psi_j(\r)
\psi_i^{\ast}(\r) \psi_i(\r') d\r d\r'
\label{hf3}
\end{eqnarray}
\noindent where $E_T^0$ is the total energy in the absence of interaction. $U(\r-\r')$ is the interaction.
 The summation $\sum_{i,j}$ is  on filled energy levels. $\sigma_i$ is
the spin of a eigenstate $\psi_i$.
We now assume that the  Coulomb interaction is   screened and that  the states $\psi_i$ are those of the
{\it non-interacting} system. This corresponds to the so-called RPA approximation.
$U(\r-\r') = U \delta(\r-\r')$ where $U=4 \pi e^2 / q_{TF}^2$, $q_{TF}$ being
 the Thomas-Fermi wave vector\cite{screening}. For such a local interaction, the Fock
term has the same structure as the Hartree term. Introducing the local density $n(\r) = \sum_{i} |\psi_i(\r)|^2$,
 one
has finally
\begin{equation}  \label{currentee}
 \langle M_{e-e} \rangle =  -\langle{\partial E_T \over \partial
B}\rangle = -{U \over
4} {\partial \over \partial B} \int \langle n^2(\r) \rangle
d\r
\end{equation}
We define the local DOS $\rho(\r,\omega)$ so that
$n(\r) = 2 \int_{-\epsilon_F}^0 \rho(\r,\omega) d\omega$ ( the factor $2$ accounts
for spin). The magnetization
can be rewritten as  \cite{Ambegaokar90,Schmid91,Montambaux95}: \begin{eqnarray} \langle M_{e-e} \rangle &=&
- U
 {\partial \over \partial B} \int \langle \rho(\r,\omega_1)
\rho(\r,\omega_2) \rangle d\r  d\omega_1  d\omega_2  \\
&=&  -{U
\over  2 \pi^2} {\partial \over \partial B} \int \langle
G^R_{\omega_1}(\r,\r)
G^A_{\omega_2}(\r,\r) \rangle d\r  d\omega_1  d\omega_2
\label{currentee2}
\end{eqnarray}

The average product in the integral is nothing but the return probabiliy defined in eq. (\ref{proba}). The interaction
contribution to
the average
magnetization can thus be written as a function of field dependent part of the return
probability \cite{Montambaux95,Argaman}:

\begin{equation}
\begin{array}{|c|}
\hline
\\
\displaystyle
\langle M_{ee} \rangle =  -{\lambda_0
 \over  \pi} {\partial \over \partial B} \int_0^\infty  P_{int}(t,B){ e^{-\gamma t} \over
t^2}   dt
\\
\\
\hline
\end{array}
 \label{currentee3} \end{equation}
where $\lambda_0=U \rho_0$ is the interaction parameter.
Considering higher corrections in the Cooper channel leads to a ladder summation
 \cite{AAZ,Eckern91,Ullmo97}, so that $\lambda_0$ should be replaced by $\lambda(t)  =
\lambda_0/(1+\lambda_0
\ln(\epsilon_F t))=1/\ln(T_0 t)$ where $T_0$ is defined as $T_0=\ep_F e^{1/\lambda_0}$ \cite{Sum}. We shall discuss later
the contribution of this renormalization.

\subsection{Typical Magnetization}

We first calculate the
typical
magnetization $M_{typ}$, defined as $M_{typ}^2 =
 \langle M^2 \rangle - \langle M \rangle^2$. From eq.(\ref{Isum}), it can
be written as:  \begin{equation}
 M^2_{typ} = 4 {\partial \over \partial B}
{\partial \over \partial B'} \int_{-\ep_F}^0 \int_{-\ep_F}^0 \ep \ep'
K(\ep-\ep',B,B') d\ep d\ep'  |_{B'=B}\label{Itypical}
\end{equation}
where $K$ is the correlation function of the DOS: $K(\ep -
\ep',B,B')=\langle \rho(\ep,B)\rho(\ep',B')\rangle - \rho_0^2$.
 $K(\varepsilon)$ has been
calculated by Altshuler
and Shklovskii \cite{Altshuler86}. A very useful semiclassical
picture has been presented by Argaman {\it et al.}, which
relates the form factor $\tilde K(t)$, the Fourier transform of
$K(\varepsilon)$,  to the integrated return probability
$P(t) = \int P(\r,\r,t) d\r $
for a diffusive particle \cite{Argaman93}:
\begin{equation}\tilde K(t) = {\Delta^2 \over 4 \pi^2} | t | P( | t | ) \label{AIS}\end{equation}

The return probability, and consequently the form factor,
is the sum of a classical and an interference term:

\begin{equation}  \label{decouplingK}
 P(t,B,B')= P_{cl}(t,{B-B'\over
2})+ P_{int}(t,{B+B'\over 2})
 \end{equation}
 Fourier transforming $K(\ep-\ep')$ and using the identity
 $\int_0^{\infty} \ep d \ep e^{i\ep t} =-1/t^2$, one obtains
straightforwardly

\begin{equation}
\begin{array}{|c|}
\hline
\\
\displaystyle
M^2_{typ}  =
\frac{1}{2\pi^2}
\int_{0} ^{+\infty}    [P''_{int}(t,B)-P''_{cl}(t,0)] {e^{-\gamma t} \over t^3} dt

\\
\\
\hline
\end{array}
 \label{Mtyp} \end{equation}
where $P''(t,B)=\partial^2 P(t,B) / \partial B^2$.

\section{Simple examples}
We have now an ensemble of quantities which are all simply written in
terms of integrals of the return probability $P(t)$.
I choose now a few examples where known results can be recovered straightforwardly with the use the general formula derived
above.
\subsection{2D electron gas}\label{2Dgas}
Consider an infinite 2D electron gas in a magnetic field. The solutions of the diffusion equation (\ref{schrodinger}) are
the Landau levels  $E_n(B) = (n+1/2) 4 e D B / \hbar$ with degeneracy $g_n=2 e B /\hbar$ so that the
return probability in a field is given by the sum (\ref{partition})\ \footnote{In the limit $B \rightarrow 0$, one recovers
the return probability for an infinite $2D$ system: $P(t)=S/(4 \pi D t)$}:
\begin{equation}
P_{int}(t,B)= {B S / \phi_0 \over \sinh{4 \pi B D t / \phi_0}}
\label{ZLandau}
\end{equation}
where $\phi_0=h/e$ is the flux quantum.
The integral  (\ref{WL})  gives the weak-localization correction in a field \cite{wl} (using the Einstein relation $\sigma= s e^2 D \rho_0$ and $\rho_0=1/(\Delta S)$):

\begin{equation}
\Delta \sigma(B)= -{ e^2 \over 2 \pi^2 \hbar} [\Psi({1 \over 2}+{\hbar \over 4 e D B \tau_e}) - \Psi({1 \over 2}+{\hbar
\over 4 e D B \tau_\phi}) ]
\label{deltasigma}
\end{equation}
where $\Psi(x)$ is the digamma function. \medskip

Consider now the  interaction contribution (\ref{currentee3}) to the weak-field magnetization. The low field behavior of
the return probability (\ref{ZLandau}) is $P(t,B)=P(t,0) +S [ -{ 2 \pi  D t \over 3 \phi_0^2} B^2+ {56
\pi^3  D^3 t^3 \over 45 \phi_0^4}B^4]$.  The second term of the expansion gives immediately the electron-electron
contribution to the susceptibility\cite{AAZ}. From eq.
(\ref{currentee3}), one has, per unit area:

\begin{equation}
\chi_{ee}= {4 \hbar D \over3 \phi_0^2} \int_{\tau_e}^{\tau_\phi} { d t \over t} {1 \over \ln T_0 t}= {4 \hbar D \over3
\phi_0^2}
\ln {\ln T_0 \tau_\phi \over \ln T_0 \tau_e}={2 \over \pi} \chi_L  k_F l_e\ln {\ln T_0 \tau_\phi \over \ln T_0 \tau_e}
\label{AsL}
\end{equation} It is larger than the $2D$ Landau suscpetibility $\chi_L=-e^2/(12 \pi m)$ by a factor $k_F l_e$.
\medskip

Using the next term of the  expansion of $P(t,B)$, one derives the variance $\delta \chi ^2$ of the susceptibility, that
one can write also in
terms of the 2D Landau susceptibility:

$${(\delta \chi^2)^{1/2} \over |\chi_L|} = \sqrt{84 \over 5 \pi} {L_\phi \over \sqrt{S}} \ k_F l_e $$

\subsection{Persistent currents in rings}

In a system where the diffusion is one-dimensional, the return probability writes $L/\sqrt{4 \pi D t}$. In a ring geometry,
there is an additional probability to reach the origin after $1,2,\cdots,m$ turns around the loop. The accumulated phase is
$4 \pi m \varphi$, where $\varphi=\phi/\phi_0$, so that the total return probability writes:

\begin{equation}
P_{int}(t,\phi)={L \over \sqrt{4 \pi  D t}}\sum_{-\infty}^\infty e^{-m^2 L^2 /4 D t} \cos(4 \pi m \varphi)
\label{Pring}
\end{equation}

By straighforward integration , this leads directly to the Fourier expansion of the average current $\langle I_{e-e} \rangle$ (\ref{currentee3}) and of the
typical current $I_{typ}$ (\ref{Mtyp}), where the current $I$ is simply given by $I=-{\partial A \over \partial \phi}=M/S$
where $S$ is the area of the ring. One gets, in the limit $L_\phi \rightarrow \infty$:

\begin{equation}
I^2_{typ}(\phi)={96 E_c^2 \over \phi_0^2} \sum_{m=1}^{\infty} {\sin^2 (2 \pi m \phi) \over m^3}
\label{couranttypique}
\end{equation}

and
\begin{equation}
\langle I_{e-e}( \phi) \rangle ={16 \lambda_0 E_c \over \phi_0} \sum_{m=1}^{\infty} {\sin 4 \pi m \phi) \over m^2}
\label{courantmoyen}
\end{equation}
as obtained for the first time in refs.( \cite{Argaman93}) and ( \cite{Ambegaokar90}) \cite{harmonics}.
For discussion of these results and comparisons to experiments see for example ref.( \cite{Gilles95}).

\subsection{Weak-localisation in cylinders}

One of the most famous experiments showing phase coherence effect on transport, is the one performed by Sharvin and Sharvin
\cite{Sharvin}
who measured the magnetoresistance of a cylinder pierced by a magnetic flux $\phi$. In this case the return probability
$P_{int}(t,\phi)$ is modulated by the flux through the cylinder. The time dependence of the return probability on the cylinder ressembles the one on the ring (eq.(\ref{Pring})), with a different power law:
\begin{equation}
P_{int}(t,\phi)={S \over {4 \pi  D t}}\sum_{-\infty}^\infty e^{-m^2 L^2 /4 D t} \cos(4 \pi m \varphi)
\label{Pcylinder}
\end{equation}
where $S$ is the area of the cylinder. By integration, one obtains immediately:
\begin{equation}
\Delta \sigma(\phi)= -{e^2 D  \over \pi \hbar} \left[ \ln {L_\phi \over l_e} + 2 \sum_{m=1}^\infty K_0(m {L \over L_\phi})
\cos(4 \pi m \varphi)\right]
\label{AAS}
\end{equation}
This expression was first obtained by Altshuler, Aronov and Spivak \cite{aas}.

\subsection{Universal conductance fluctuations in a quasi-1D wire}

Consider a quasi-1D wire of length $L$ connected to leads. The return probability in this case is
$P(t) = \sum_q e^{-D q^2 t}$
where the modes are quantized as $q=n \pi / L$, with $n >0$. Eq.(\ref{UCF}) gives, (using $\sigma=s e^2 D \rho_0=s e^2 D/(\Delta L)$)
\begin{equation}
\langle \delta \sigma^2 \rangle  =  \sigma^2 {3  \Delta^2 \over \beta \pi^2 \hbar^2}\sum_{q\neq 0}{1 \over D^2 q^4}=
{12 s^2\over \beta \pi^4} L^2  \left({e^2 \over h}\right)^2 \sum_{n>0} {1 \over n^4}
 \label{UCF2} \end{equation}
so that the fluctuation of the conductance $G=\sigma/L$ is universal \cite{Altshuler86}:
$$\langle \delta G^2 \rangle = {2 s^2\over 15 \beta }$$
\section{Temperature effect}

The above results which have been obtained for $T=0K$ . Thermal broadening can be taken into account straightforwardly.
Let us take the example of the typical magnetization. In eq. (\ref{Itypical}), thermal functions must be inserted so that it has now the form:

$$
\int \int d\ep  d\ep' K(\ep-\ep')
F(\ep) F(\ep')$$ where  $F(\ep)={1 \over \beta} \ln (1 + e^{\beta(\ep_F-\ep)})$ is the integral of the Fermi factor.
By Fourier transform, this integral is simply rewritten as $$ \int  {\pi^2 T^2 \over ( t \sinh \pi T t )^2} \tilde{K}(t) dt
$$
so that eq. (\ref{Mtyp}) becomes:
\begin{equation}
M^2_{typ}  =
\frac{1}{2\pi^2}
\int_{0} ^{+\infty}    [P''_{int}(t,B)-P''_{cl}(t,0)] {\pi^2 T^2 t \over ( t \sinh \pi T t )^2}e^{-\gamma t} dt
 \label{MtypT} \end{equation}

Let us define a temperature characteristic length $L_T$ by $L_T^2= D /T$ ($\hbar=k_B=1$). In the limit $L_T < L_\phi$, using the field expansion of $P(t,B)$ written in section \ref{2Dgas}, eq.(\ref{MtypT}) becomes:
\begin{equation}
\delta \chi^2(T)=
{112 \pi \over 5} S {D^3 \over \phi_0^4}
\int_{0} ^{+\infty}   \left({\pi T t \over \sinh \pi T t}\right)^2 dt
\end{equation}where $\chi=\partial M /\partial B$.
The integral is $\pi/(6 T)$ so that
$${(\delta \chi^2)^{1/2} \over |\chi_L|} = \sqrt{14 \over 5 } {L_T \over \sqrt{S}} \ k_F l_e $$
as obtained in a different way by Raveh {\it et al.} \cite{Shapiro}.
Other temperature dependences of transport and thermodynamic coefficients can be obtained in the same way.

\section{Diffusion on Graphs}

The calculation of the above quantities can be extended to the case of any structure -- called a network -- made of quasi-
one-dimensional wires.
First, we note that the quantities of interest have all the same structure:

\begin{equation} \int t^\alpha P(t) e^{-\gamma t} dt
\end{equation}
 where $P(t)=\sum_n e^{- E_n t}$. The time integral of $P(t)$ can be straightforwardly written in terms of a quantity
called the spectral determinant $S_d(\gamma)$:
\begin{equation}
{\cal P} \equiv
  \int_0^\infty dt  P(t)
 = \sum_n \frac{1}{E_n + \gamma}
= \frac{\partial}{\partial \gamma} \ln {\cal S}_d (\gamma)
\label{diff.spectral}
\end{equation}
where $S_d(\gamma)$ is, within a multiplicative constant independant of $\gamma$:
\begin{equation}
{\cal S}_d (\gamma) = \prod_n (\gamma + E_n)
\label{SD}
\end{equation}
 $E_n$ being the eigenvalues of the diffusion equations (\ref{diff1}) or {\ref{diff2}).
Using standard properties of Laplace transforms, the above time integrals can be rewritten in terms of the spectral
determinant, so that the  physical quantities described above  read \cite{Pascaud99}:
\begin{eqnarray}
\Delta \sigma &=&                  - \sigma   \ {\Delta \over \pi}\ \ \ \          {\partial \over \partial \gamma} \ln
S_d(\gamma)\\
\langle \delta \sigma^2 \rangle &=& - \sigma^2 \ {3 \Delta^2 \over \beta \pi^2}\ \  {\partial^2 \over \partial
\gamma^2} \ln
S_d(\gamma)\\
M^2_{typ}  & = &                     \frac{1}{2\pi^2}\ \ \
\int_\gamma ^{+\infty} d\gamma_1
(\gamma - \gamma_1)\frac{\partial ^2}{\partial B ^2} \ln {\cal S}_d (\gamma_1) \left | _{0} ^{B}
\right .
\label{Mtyp2}
\\
\langle M _{ee} \rangle  & =  &
 \frac{\lambda_0}{\pi}\ \ \
\int_\gamma ^{+\infty} d\gamma_1
\frac{\partial}{\partial B}  \ln {\cal S}_d (\gamma_1)
\label{Mee}
\end{eqnarray}
These expressions are quite general, strictly equivalent to expressions (\ref{WL},\ref{UCF},\ref{currentee3} and
\ref{Mtyp}).
In the case of a ring or a graph geometry, the integral converges at the upper limit. For the case
of a magnetic field in a
bulk system, this limit should be taken as $1/\tau_e$ where $\tau_e$ is the elastic time.
The problem now remains to calculate the spectral determinant on  graphs.

 By solving the diffusion equation on
each link, and then imposing
Kirchoff type conditions on the nodes of the graph,  the problem can be  reduced to the solution of a
system of $N$ linear
equations relating the eigenvalues at the $N$ nodes. Let us introduce the $N \times N$ matrix
$M$ \cite{Doucot86,Sgraphs}:
\begin{equation}
M_{\alpha \alpha} = \sum_\beta \coth(\eta_{\alpha \beta})
\ \ , \ \
M_{\alpha \beta} = - \frac{e^{i \theta_{\alpha \beta}}}
{\sinh \eta_{\alpha \beta}}
\label{matrixM}
\end{equation}
The sum $\sum_\beta$ extends to all the nodes $\beta$ connected to the node $\alpha$; $l_{\alpha
\beta}$ is the length of
the link between $\alpha$ and $\beta$. $\eta_{\alpha \beta} = l_{\alpha
\beta}/L_\phi$.
The off-diagonal coefficient $M_{\alpha \beta}$ is non zero only if there is a link connecting the
nodes $\alpha$ and
$\beta$.
$\theta_{\alpha \beta} = (4 \pi /\phi_0)\int_\alpha ^\beta A.dl$ is the circulation of the vector
potential
between $\alpha$ and $\beta$.
$N_B$ is the number of links in the graph.
It can then be shown that the integrated return probability can be
rewritten as: $
{\cal P} =
\frac{\partial}{\partial \gamma}
\ln
{\cal S} _d $ where the spectral determinant ${\cal S} _d$ is given by \cite{Pascaud99}:

\begin{equation}
{\cal S} _d =
\left ( \frac{L_\phi}{L_0} \right ) ^{N_B -N} \
\prod _{(\alpha \beta)} \sinh \eta_{\alpha \beta} \ \det M
\label{det.spectral.Q1D} \end{equation}

 $L_0$ is an arbitrary length independent of $\gamma$ (or $L_\phi$).
We have thus transformed the spectral determinant which is an infinite product in a finite product
related to $\det M$.

\medskip

Let us come back to a   network made of diffusive connected rings.
Experimentally,  the coherence length is of the order of the perimeter of one ring so that only a
few harmonics of the flux
dependence may be observed.  It is then useful to make a perturbative expansion. We split the
matrix as $M=D-N$, where $D$ is a diagonal matrix:
$D_{\alpha \alpha}=M_{\alpha \alpha} \approx z_\alpha $ to the lowest order in
$L_\phi$ ($z_\alpha$ is the connectivity of the node $\alpha$);
$N_{\alpha \beta}=M_{\alpha \beta}
\approx 2 e^{-l_{\alpha \beta}/L_\phi} e^{i \theta_{\alpha \beta}}$. Expanding $\ln \det (I -D^{-
1} N)=\mbox{Tr}
[\ln (I -D^{-1} N)]$, we
have:
\begin{equation}
\ln \det M
=\ln \det D -
\sum_{n \geq 1} \frac{1}{n} \mbox{Tr} [ (D^{-1} N )^n]
\end{equation}
We call ``loop'' $l$, a set of $n$ nodes linked by $n$ wires  in a closed loop. The length $L_{l}$
of a loop $l$ is the sum
of the lengths of the $n(l)$ links. The flux dependent part of $\ln {\cal S}$ can be expanded as:
\begin{equation}
\ln {\cal S} = -2 \sum_{\{ l \}}
\frac{2}{z_1} \ldots \frac{2}{z_{n(l)}} e^{-L_{l}/L_\phi} \ \cos(4\pi \phi_{l} / \phi_0)
\label{dev.1erordre}
\end{equation}
$\phi_{l}$ is the flux enclosed by the loop $l$.

For example, we
consider the different configurations of connected rings shown on Table.\ref{tableau.facteur.geometrique}. The first
harmonics of the total magnetization, to the first order in
$\lambda_0$ is:
\begin{equation}
\langle M _{ee} \rangle = 2 G { \lambda_0 e D \over \pi^2} (L/L_\phi+1)  e^{-L/L_\phi}
\end{equation}
where $G$ is shown in Table. (\ref{tableau.facteur.geometrique}) \cite{harmonics}.

\begin{table}[!Ht]\label{tableau.facteur.geometrique}
\begin{center}
$
\begin{array}{|c|c|c|c|c|c|c|c|}
\hline
 & & & & & &&
\\
\epsfysize 0.5cm
\epsffile{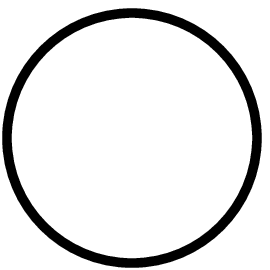}
&
\epsfysize 0.5cm
\epsffile{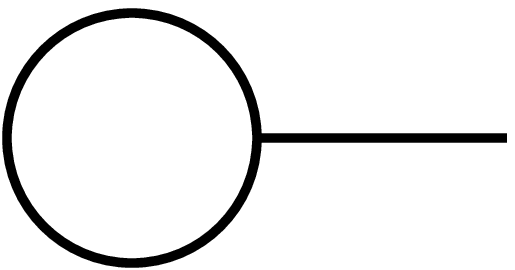}
&
\epsfysize 0.5cm
\epsffile{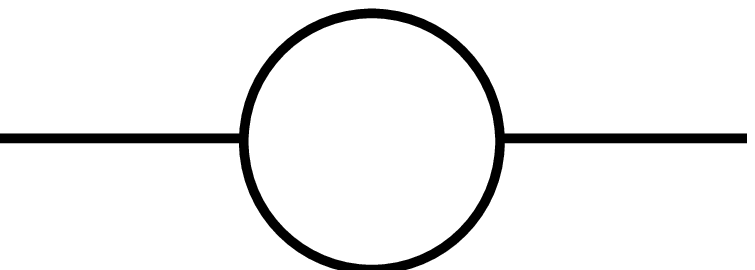}
&
\epsfysize 0.7cm
\epsffile{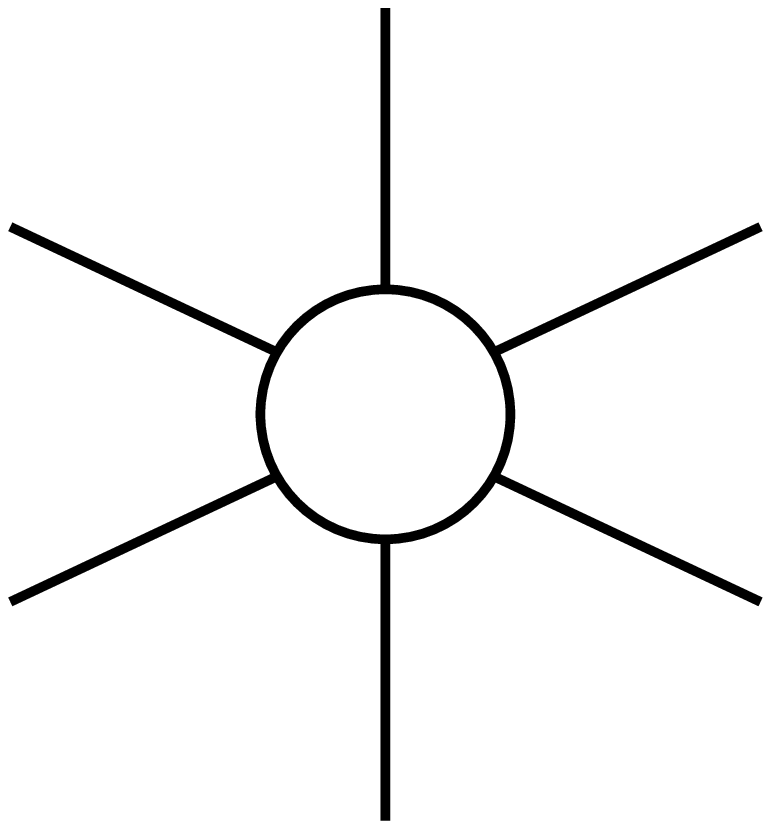}
&
\epsfysize 0.5cm
\epsffile{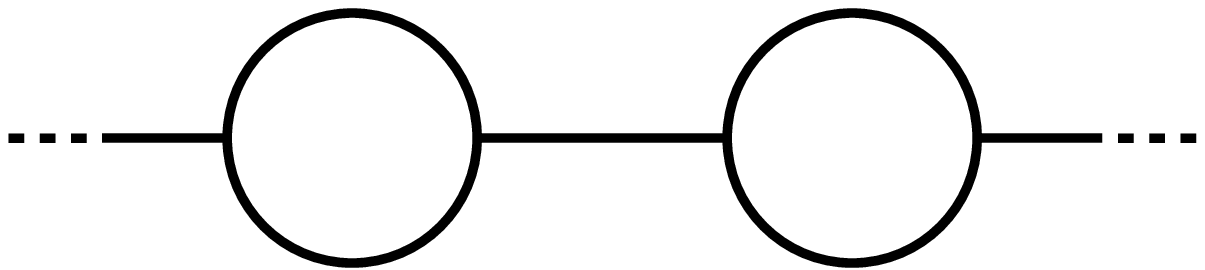}
&
\epsfysize 0.5cm
\epsffile{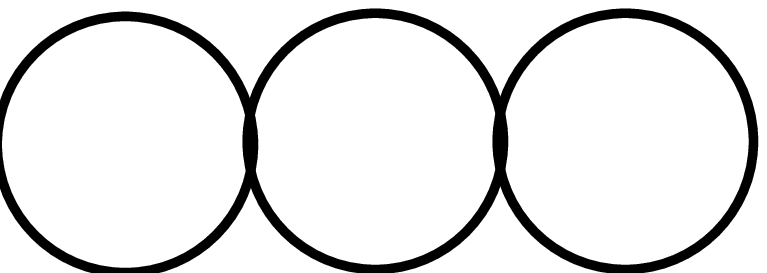}
&
\epsfysize 0.5cm
\epsffile{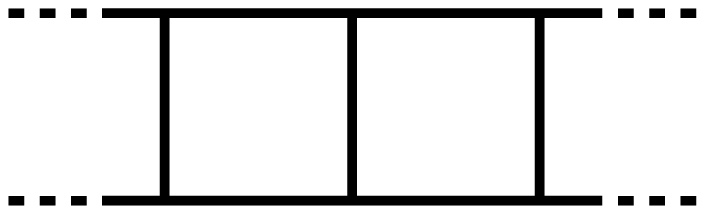}
&
\epsfysize 0.7cm
\epsffile{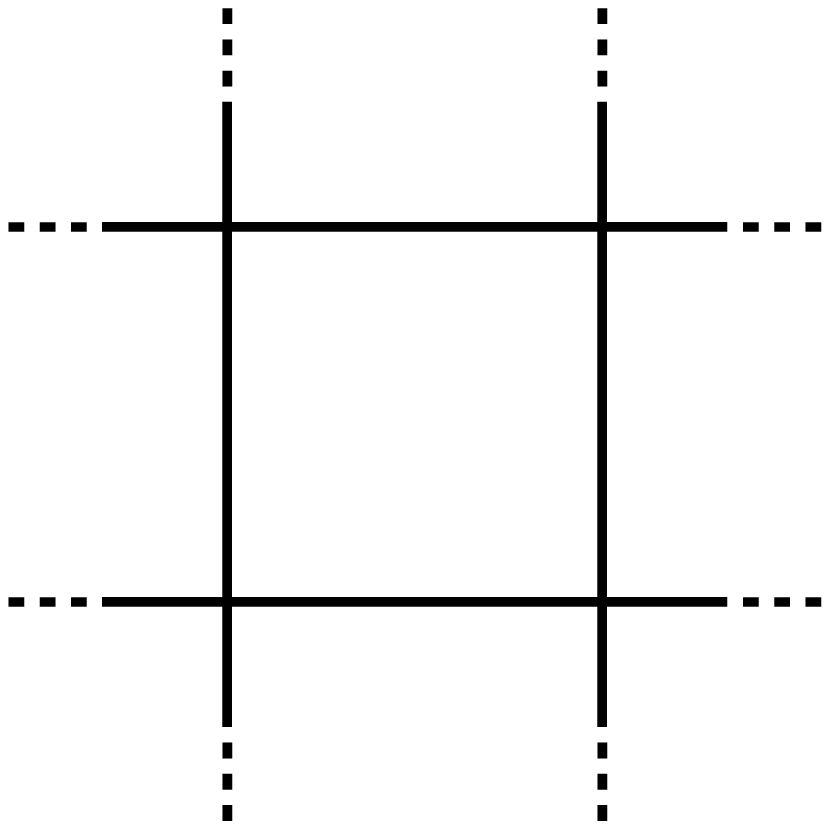}
\\
& & & & & & &
\\
\hline
& & & & & & &
\\
1 & {2 \over 3} & {4 \over 9} & \left({2 \over 3}\right)^n &  {4 \over 9}
& {1 \over 4} & {16 \over 81} &{1 \over 16}
\\
& & & & & & &
\\
\hline
 \end{array}
$
\caption{The geometrical factor $G$ for different networks}
\end{center}
\end{table}

\section{Mapping}\label{map}

We now  wish to emphasize
an interesting
correspondence  between the HF magnetization of a phase coherent {\it interacting diffusive}
system
and the grand canonical magnetization $M_0$ of the corresponding {\it non-interacting clean} system. The
latter can also be
written  in term of a spectral determinant. The grand canonical magnetization $M_0$ is given quite generally by:
\begin{equation}
M_{0}  =
-\frac{\partial \Omega}{\partial B}= - \frac{\partial }{\partial B}\int_{0} ^{\epsilon_F}
d\epsilon N(\epsilon)
\label{M1}
\end{equation}
where the integrated DOS is
\begin{equation}
N(\epsilon)  =
 -\frac{1}{\pi} \mbox{Im}
\sum _{\epsilon_\mu} \ln (\epsilon_\mu - \epsilon_+)=-\frac{1}{\pi} \mbox{Im}
 \ln {\cal S} (\epsilon_+)
\label{M2}
\end{equation}
 where $\epsilon_{+}=\epsilon+i0$, ${\cal S}(\epsilon)= \prod_{\epsilon_\mu} b_\mu(\epsilon_\mu -
\epsilon )={\cal S}_d
(\gamma=-
\epsilon)$.
where $\epsilon_\mu$ are the eigenvalues of the Schr\"odinger
equation. For a clean system these eigenvalues are the same as
those of the diffusion equation, with the substitutions $D
\rightarrow \hbar/(2 m)$ and $2e \rightarrow e$ \cite{Boundaries}.

Comparing eqs.(\ref{M1},\ref{M2}) with eq.(\ref{Mee}), we can now formally relate $M_0$  and the HF magnetization $\langle
M_{ee}\rangle$ of the same diffusive system:

\begin{equation}
M_{0} = - \mbox{lim}_{\lambda_0 \rightarrow 0} {1 \over \lambda_0} \mbox{Im}
[\langle M_{ee} \rangle (-\epsilon_F -i0) ]
\label{mapping}
\end{equation}
As a simple illustration, consider the orbital magnetic
susceptibility of an infinite disordered plane. For a disordered
conductor, it is given by eq. (\ref{AsL})
 \cite{AAZ}:
After replacing $\gamma$  by $-\epsilon_F-i0$,
taking the
imaginary part of the
logarithm and replacing $D$ and $2e$, we recover  the Landau
susceptibility for the clean
system:
$\chi_{0} = -e^2/(24\pi  m)$.
\medskip

\section{Conclusion}
In conclusion, we have shown how to relate phase coherent transport and thermodynamic properties to the return probability for a diffusive particule. It is then possible to calculate straightforwardly these quantites by simple integrals of this return probability in simple geometries. For networks made of diffusive wires, we have developed a formalism which relates {\it directly} the persistent current,
and the transport properties   to the determinant of a
matrix  describing the connectivity of the graph. From a loop expansion of this determinant, simple predictions for the
persistent current in any geometry can now be compared with forthcoming
experiments on connected and disconnected rings. We have also found a correspondence between the phase coherent
contribution to the orbital magnetism of a disordered
interacting system and the orbital response of the corresponding clean non-interacting system.
\bigskip

\section*{Acknowledgments}
The formalism of the chapter \ref{map} has  been developed in collaboration with M. Pascaud.
I would like to thank Professors J. Tr\^an Thanh V\^an and Nguyen Van Hieu, for the remarkable organisation of this conference.


\begin{thebibliography}{99}
\bibitem{Boundaries} The boundary conditions may be different: for the isolated system trapped
within infinite well, the Schr\"odinger equation assumes Dirichlet
boundary conditions while the diffusion assumes the condition of
zero current.
\bibitem{canonical} If the number of particle is fixed in each ring (so that $\epsilon_F$ has to be flux dependent), there
is indeed a small contribution to the average persistent current, called the "canonical contribution". It is smaller that
the contribution of the interactions discussed in the next paragraph. For a review see ref. \cite{Gilles95}
\bibitem{screening}We assume perfectly screened interaction for simplicity.
Conclusions are not affected by this hypothesis. More generally,
$U$ is a function of $k_F/\kappa$ where $\kappa$ is the
Thomas-Fermi vector\cite{Ambegaokar90}.
\bibitem{harmonics}For an harmonic $p$ of the magnetization, corresponding to a winding number $m$ in the diffusion process, one should
renormalize the interaction parameter
because of the Cooper
renormalization $\lambda=\lambda_0/(1+\lambda_0 \ln \epsilon_F/(\pi E_c/m^2))$ \cite{Eckern91}.
\end{thebibliography}
\end{document}